\documentclass[sigplan,10pt]{acmart}
\usepackage{hyperref}
\usepackage{graphicx}
\usepackage{caption}
\usepackage{adjustbox}
\usepackage{mathtools}
\usepackage{amsmath}
\usepackage{paralist}
\usepackage{multirow}
\usepackage{ulem}
\usepackage[ruled, linesnumbered, noend]{algorithm2e} 
\usepackage[labelformat=simple]{subcaption}
\usepackage{enumitem}
\usepackage{xcolor}

\setlist[itemize]{leftmargin=*}

\begin{document}

\acmYear{2026}\copyrightyear{2026}
\setcopyright{cc}
\setcctype[4.0]{by}
\acmConference[EUROSYS '26]{European Conference on Computer Systems}{April 27--30, 2026}{Edinburgh, Scotland Uk}
\acmBooktitle{European Conference on Computer Systems (EUROSYS '26), April 27--30, 2026, Edinburgh, Scotland Uk}
\acmDOI{10.1145/3767295.3769382}
\acmISBN{979-8-4007-2212-7/26/04}



\graphicspath{ {./figures/} }
\title{Scaling LLM Test-Time Compute with Mobile NPU on Smartphones}

\author{Zixu Hao}
\affiliation{
    \institution{Tsinghua University}
    \country{}
}
\email{haozx23@mails.tsinghua.edu.cn}

\author{Jianyu Wei}
\affiliation{
    \institution{University of Science and Technology of China}
    \country{}
}
\email{noob@mail.ustc.edu.cn}

\author{Tuowei Wang}
\affiliation{
    \institution{Tsinghua Univeristy}
    \country{}
}
\email{wtw23@mails.tsinghua.edu.cn}

\author{Minxing Huang}
\affiliation{
    \institution{Tsinghua University}
    \country{}
}
\email{huangmx25@mails.tsinghua.edu.cn}

\author{Huiqiang Jiang}
\affiliation{
    \institution{Microsoft Research}
    \country{}
}
\email{hjiang@microsoft.com}

\author{Shiqi Jiang}
\affiliation{
    \institution{Microsoft Research}
    \country{}
}
\email{shijiang@microsoft.com}

\author{Ting Cao}
\authornote{Corresponding authors.}
\affiliation{
    \institution{Institute for AI Industry Research (AIR), Tsinghua University}
    \country{}
}
\email{tingcao@mail.tsinghua.edu.cn}

\author{Ju Ren}
\authornotemark[1]
\affiliation{
    \institution{Tsinghua University}
    \country{}
}
\email{renju@tsinghua.edu.cn}

\renewcommand{\shortauthors}{Zixu Hao et al.}

\cfoot{}

\begin{abstract}
Deploying Large Language Models (LLMs) on mobile devices faces the challenge of insufficient performance in smaller models and excessive resource consumption in larger ones. This paper highlights that mobile Neural Processing Units (NPUs) have underutilized computational resources, particularly their matrix multiplication units, during typical LLM inference. To leverage this wasted compute capacity, we propose applying parallel test-time scaling techniques on mobile NPUs to enhance the performance of smaller LLMs. However, this approach confronts inherent NPU challenges, including inadequate hardware support for fine-grained quantization and low efficiency in general-purpose computations. To overcome these, we introduce two key techniques: a hardware-aware tile quantization scheme that aligns group quantization with NPU memory access patterns, and efficient LUT-based replacements for complex operations such as Softmax and dequantization. We design and implement an end-to-end inference system that leverages the NPU’s compute capability to support test-time scaling on Qualcomm Snapdragon platforms. Experiments show our approach brings significant speedups: up to 19.0× for mixed-precision GEMM and 2.2× for Softmax. More importantly, we demonstrate that smaller models using test-time scaling can match or exceed the accuracy of larger models, achieving a new performance-cost Pareto frontier. 

\end{abstract}
\maketitle
\thispagestyle{empty}

\section{Introduction}


With the advancements of commodity hardware and algorithms, deploying Large Language Models (LLMs) on mobile devices is becoming increasingly feasible. Many language models tailored for mobile devices have emerged, including Llama 3.2~\cite{llama32}, MiniCPM~\cite{minicpm,minicpm-v}, Gemma~\cite{gemma1,gemma3}. However, these models generally underperform compared to their larger counterparts. A straightforward approach to improve model performance is to scale up the model size, yet this significantly increases memory consumption and bandwidth requirements, posing serious challenges for resource-constrained mobile platforms.


Recently, a new paradigm named \textbf{test-time scaling} have introduced new opportunities to enhance LLM capabilities through increased inference-time computation. Parallel test-time scaling methods involve generating multiple paths and selecting the best sample among a number of generation candidates~\cite{cot-sc,wu2024inference,snell2024scaling,huggingface2025testtime,brown2024large,liu2025can1bsurpass405b,tts-survey,chen2025sets}. So far, these methods are limited to cloud or offline settings where computational resources are abundant.

Intuitively, employing test-time scaling techniques to enhance LLM's generation quality on mobile devices may seem impractical. Mobile devices such as smartphones are typically considered resource-constrained, while LLM inference is known for its high resource consumption. On top of this, scaling compute resources at runtime requires even more computation.

However, recent integration of Neural Processing Units (NPUs) in mobile SoCs has begun to shift this landscape. Vendors including Qualcomm, Intel, and AMD have designed and integrated NPUs to accelerate AI workloads~\cite{hexagon,intel-npu,amd-xdna}. These NPUs not only achieve high peak computing power but also undergo rapid evolution: Qualcomm claims that its Hexagon NPU in Snapdragon X Elite delivers 45 TOPS of INT8 performance~\cite{qualcomm_on_device_ai}, while recent generations of AMD NPUs have achieved $3.1\times$\footnote{The value is obtained by dividing the 50 TOPS of the AMD Ryzen AI 9 HX 370 NPU by the 16 TOPS of the AMD Ryzen 7 8845HS NPU.} performance improvements~\cite{amd_processors_specs}. These developments are transforming the computation capabilities of mobile devices.



We discover that mobile NPUs achieve high peak performance through dedicated matrix multiplication units that operate on large matrix tiles. However, in typical LLM inference, GEMM operations often degenerate into GEMV during the decoding phase, resulting in low hardware utilization and waste of computing capabilities of the large-tile optimized matrix units. This underutilization presents an opportunity: test-time scaling methods that increase sampling parallelism can leverage this available compute capacity without substantially adding to inference overhead.

Despite this potential, achieving efficient test-time scaling with mobile NPUs faces significant hardware challenges, which we categorize into two aspects:

\textbf{Precision:} Mobile NPUs were originally designed for coarse-grained quantized models and lack native support for fine-grained group quantization, which is essential for modern LLMs deployed in low bits. We observe that models quantized with conventional per-channel methods suffer severe performance degradation on reasoning tasks that are critical for test-time scaling.

\textbf{Efficiency:} While NPUs excel at matrix multiplication, their general-purpose vector units offer limited compute throughput and memory bandwidth. Many key non-matrix computations in LLM inference for test-time scaling must run on vector units, becoming a prominent bottleneck. Furthermore, the mismatch between the wide SIMD vector components and data granularity, coupled with the hardware's memory access limitations, makes it difficult for software to fully utilize the computing power of vector units, further exacerbating the problem. 

To address these challenges, we present an end-to-end LLM inference system that leverages the abundant compute capacity of mobile NPUs to support test-time scaling workloads. To meet on-device resource constraints and precision requirements, we mainly adopt weight-only 4-bit fine-grained group quantization. For the resulting efficiency challenges, our solution incorporates the following key techniques:


\textbf{Hardware-aware Tile Quantization Scheme:} We present a novel matrix and vector unit-aware quantization layout. Through weight layout transformations before and after quantization, we apply fine-grained group quantization on hardware-friendly tiles and align with NPU's memory access patterns, thereby minimizing runtime memory access overhead and maximizing vector compute utilization.

\textbf{Efficient LUT-Based Computation:} We replace complex key operations, including exponential computation in Softmax and the dequantization process in mixed precision GEMM, with efficient table lookup (LUT) instructions, alleviating computation bottleneck on the vector units.

We evaluate our system across three generations of Qualcomm Snapdragon platforms. Our proposed techniques bring up to $19.0\times$ speedup for mixed-precision GEMM and $2.2\times$ acceleration for Softmax compared to baselines, respectively. The results demonstrate the effectiveness of exploiting mobile NPUs for LLM test-time scaling workloads. Notably, we show that test-time scaling achieves state-of-the-art performance-cost trade-offs: using test-time scaling with smaller models can match or even surpass the performance of larger models running without scaling. To the best of our knowledge, this is the first work to explore the feasibility and evaluate the trade-offs of test-time scaling methods for LLMs with NPUs on mobile devices. Our contributions are summarized as follows:


\begin{itemize}
    \item We analyze the architecture of modern mobile NPUs and identify underutilization of the specialized matrix units during the LLM decoding phases.
    \item We present two techniques: a hardware-aware tile quantization scheme and LUT-based computations to accelerate LLM test-time scaling on mobile NPUs.
    \item We design and implement an end-to-end LLM inference system\footnote{Our code is available at https://github.com/haozixu/llama.cpp-npu (main repo) and https://github.com/haozixu/htp-ops-lib (op library)} that leverages mobile NPUs to support test-time scaling workloads with minimal dependency on proprietary software stacks.
    \item We demonstrate that test-time scaling can effectively leverage otherwise wasted NPU compute capacity to enhance the generation quality of on-device small language models, achieving Pareto-frontier performance in accuracy and cost compared to traditional model scaling. It opens up new opportunities for deploying LLMs on mobile devices.
\end{itemize}

\section{Background}

\subsection{Scaling LLM Computation at Test-Time}

\begin{figure}
    \centering
    \includegraphics[width=0.8\linewidth]{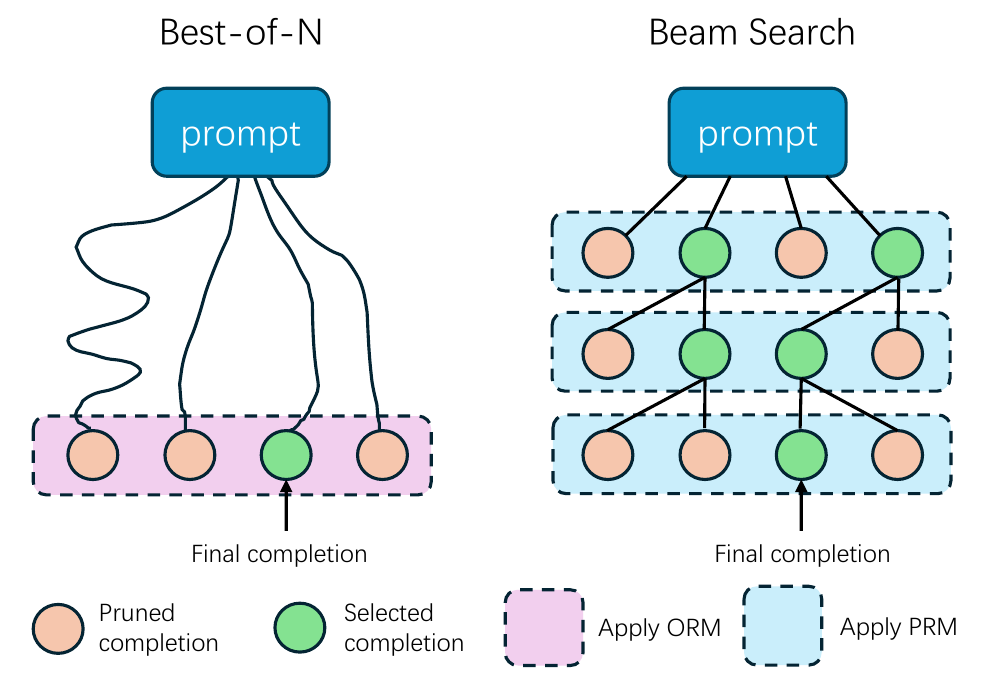}
    \caption{Two typical test-time scaling methods: Best-of-N and Beam Search.}
    \label{fig:tts-methods}
\end{figure}

Parallel test-time scaling emerges as a popular and effective new paradigm to improve model accuracy without modifying model parameters; instead, it devotes more computation at test-time. The simplest test-time scaling methods are majority-voting and self-consistency~\cite{cot-sc,brown2024large}, which are used to select the most consistent answer from multiple sets of generated samples. For math or programming problems with verifiable outcomes and domains with reward models (i.e., Outcome Reward Models), the highest scoring option can be chosen from completed sample sets, a strategy termed Best-of-N~\cite{snell2024scaling}. Through lookahead rollouts, methods similar to Monte Carlo Tree Search (MCTS) can select optimal paths from partially generated sequences, leading to the derivation of Process Reward Models (PRMs)~\cite{math-shepherd,processbench,Skywork-o1,dwivedi2023making} that directly score intermediate results. With PRM assistance, lookahead-free step-level Beam Search~\cite{snell2024scaling,wu2024inference} dynamically discards low-quality generation paths to balance exploration and exploitation. Figure~\ref{fig:tts-methods} illustrates the algorithm of two popular test-time scaling methods.

\subsection{Neural Processing Units}

With the growth of AI workloads, modern SoCs are increasingly integrating NPUs to accelerate neural network inference~\cite{hexagon,amd-xdna,intel-npu}. NPUs feature specialized acceleration of low-precision, computationally intensive core neural network operations (e.g., GEMM), delivering extremely high computational throughput while maintaining good power efficiency.

\begin{figure}[h]
    \centering
    \includegraphics[width=0.8\linewidth]{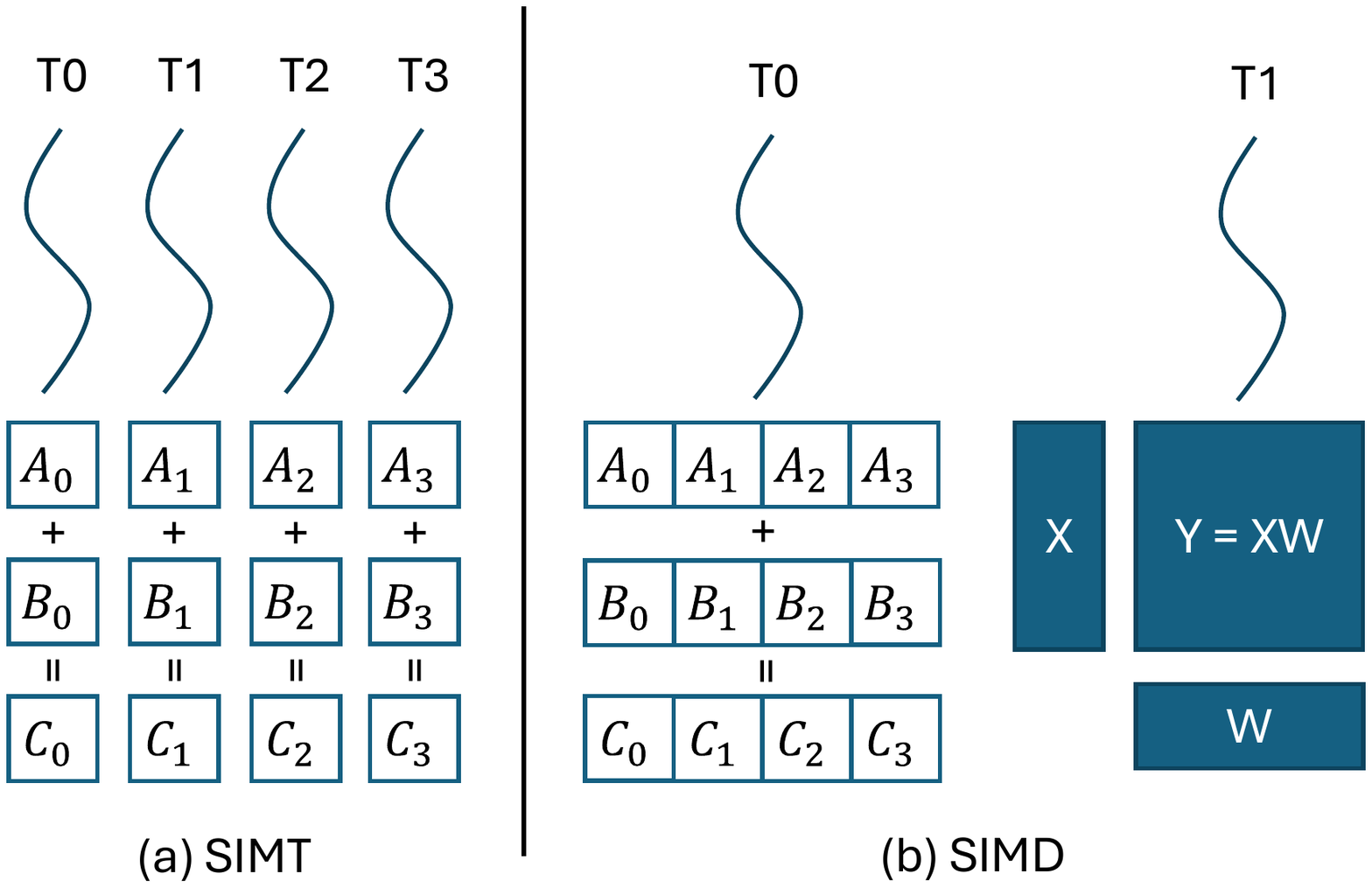}
    \caption{Comparison of (a) GPU's SIMT execution model and (b) NPU's SIMD execution model.}
    \label{fig:hw-execution-model}
\end{figure}

A widely adopted NPU architecture employs a "vector + matrix" combination, where the matrix unit accelerates operations like matrix multiplication and convolution, and the vector unit handles general-purpose computations such as normalization and complex activation functions. Well-known examples including Qualcomm’s Hexagon NPU~\cite{hexagon}, Huawei’s Ascend NPU~\cite{huawei-ascend}, AMD's XDNA NPU~\cite{amd-xdna}, Intel NPU~\cite{intel-npu}, and Intel’s Gaudi HPU~\cite{intel-gaudi3} all utilize this type of architecture. Such NPUs differ significantly from common GPUs in their hardware execution model. As shown in Figure~\ref{fig:hw-execution-model}, in the GPU’s SIMT model, different threads can independently perform branching, memory access, and computation, whereas in the NPU’s SIMD-based execution model, a single thread operates on large vector or matrix data blocks. At the hardware level, NPUs typically employ fewer hardware threads and use VLIW architectures to reduce control logic overhead. Compared to GPUs, NPUs sacrifice programming flexibility and ease-of-use in exchange for higher execution efficiency and energy efficiency.

\section{Motivation and Challenges}

In this section, we first introduce some key features of mobile NPUs, and then analyze the opportunities of leveraging NPUs' free compute as well as the challenges in implementing efficient system for test-time scaling workloads.

\subsection{Qualcomm's Hexagon NPU} \label{sec:hexagon-npu}


The Hexagon NPU on Qualcomm’s Snapdragon SoC is a representative mobile NPU due to its typical architecture, widespread adoption, and relatively accessible SDK. Therefore, we use it to demonstrate the core features of mobile NPUs.

\subsubsection{Programming Interface}

The primary approach to program Qualcomm's Hexagon NPU is through Qualcomm AI Engine Direct~\cite{qualcomm-qnn} (often referred to as QNN), a proprietary, closed-source DNN inference framework. In most cases, developers cannot customize high-performance low-level kernels even though the full LLVM toolchain for Hexagon NPU is provided in the Hexagon SDK, mainly because the instructions for the matrix unit remain undisclosed. We are able to utilize the FP16 matrix unit by reverse engineering the undocumented instructions in the binary libraries.

\subsubsection{Architecture}

\begin{figure}[h]
    \centering
    \includegraphics[width=1.0\linewidth]{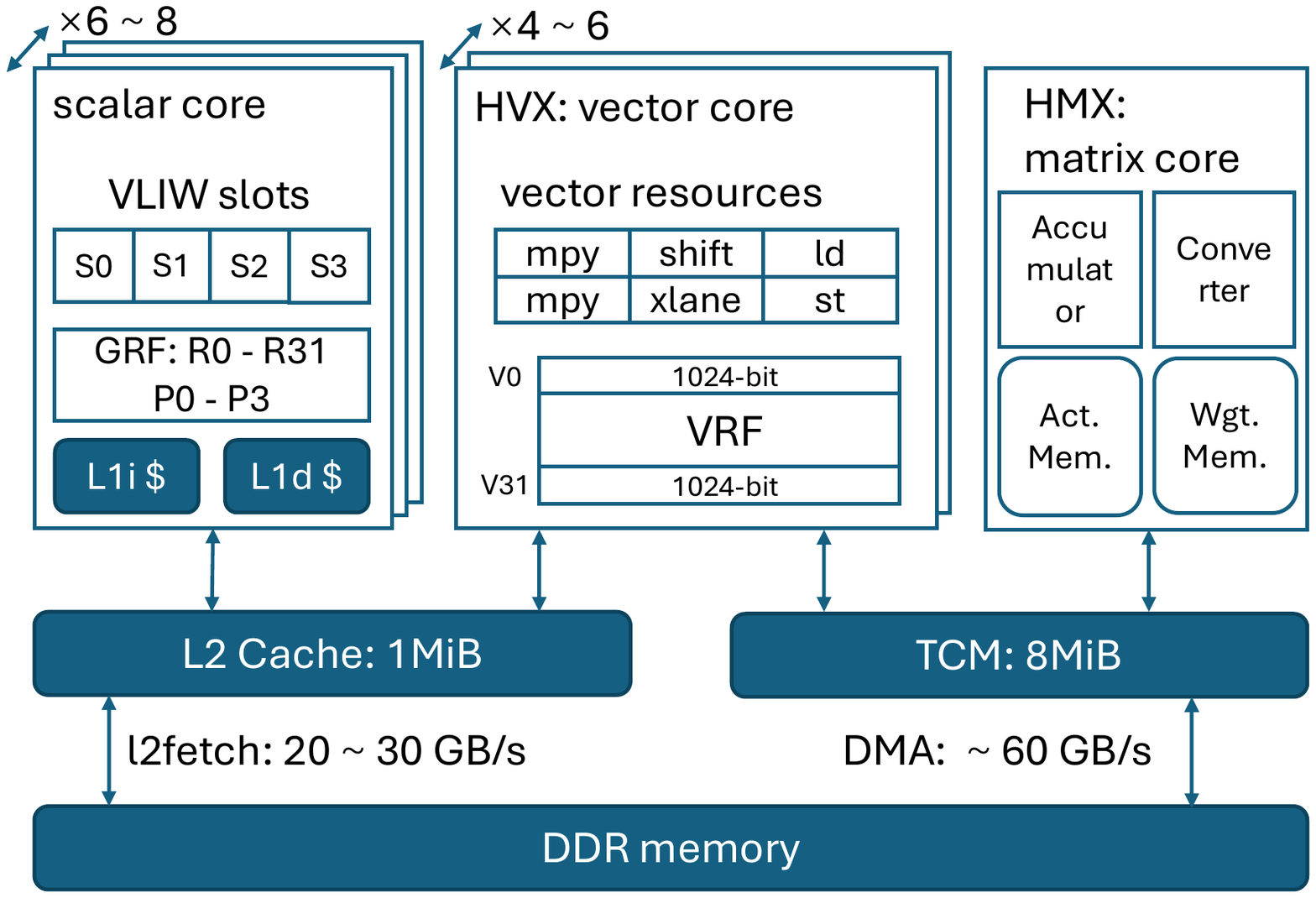}
    \caption{Hexagon NPU Architecture.}
    \label{fig:hexagon-npu-arch}
\end{figure}

\paragraph{Computation Units.} The Hexagon NPU features a typical hybrid architecture of “vector + matrix”. Its vector and matrix units are named HVX (Hexagon Vector eXtension) and HMX (Hexagon Matrix eXtension), respectively. The Hexagon NPU incorporates 6 to 8 scalar VLIW hardware threads for logical control. All vector or matrix instructions are issued from one of the four VLIW slots in a scalar core. The HVX unit context comprises 32 vector registers with a width of 1024 bits, and the number of such units ranges from 4 to 6. The number of HMX units is deduced to be 1 or 2.

\paragraph{Memory Subsystem.} The Hexagon NPU includes a shared 1 MiB L2 cache and 8 MiB of TCM (Tightly Coupled Memory), the latter being a segment of software-managed on-chip memory. The HVX can read data from either the L2 cache or the TCM. Vector scatter/gather operations and all HMX instructions can only access TCM. Data can be loaded from DDR memory into the L2 cache and TCM via the \texttt{l2fetch} instruction and DMA mechanisms, respectively. Both support asynchronous transfers of 1D or 2D tensor data.

\paragraph{The HMX Unit.} The powerful matrix multiplication capabilities of the Hexagon NPU originate from the HMX component. According to Qualcomm, the HMX unit supports various precisions, including INT4, INT8, INT16, and FP16~\cite{hexagon}. The following introduction is based mainly on FP16 HMX, with relevant information derived from reverse engineering, the Hexagon SDK, the QNN SDK, and publicly available information from Qualcomm.


The basic data unit for HMX operations is a tile, where each tile contains a matrix of a specific size. For FP16 HMX, a tile measures 32*32, occupying 2 KiB of space. The HMX unit can load several tiles of weight memory and activation memory from the TCM. After performing matrix multiplication on each pair of matrix tiles, it accumulates the results into an internal accumulator. Finally, it outputs a tile corresponding to the accumulator. Meanwhile, the HMX unit can independently scale and add biases to each channel (column) of the output tile.

\begin{figure}
    \centering
    \includegraphics[width=0.9\linewidth]{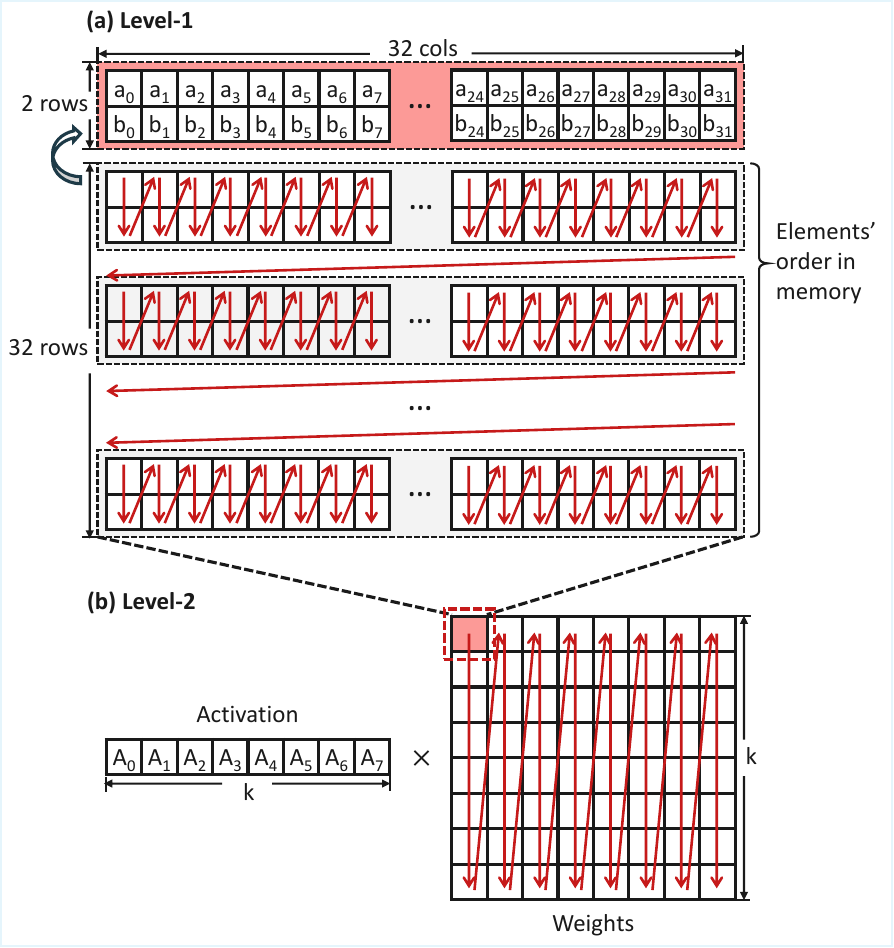}
    \caption{(a) The memory layout of FP16 HMX tile. Each tile corresponds to a 32 * 32 matrix and takes up 2048 bytes. Every two rows are permuted, having the same layout as the transposed 2 * 32 sub-matrix. (b) The overall memory layout for HMX-based GEMM. The weight tiles are arranged in column-major layout since the hardware performs inner-product at tile level.}
    \label{fig:hmx-fp16-crouton-tile-layout}
\end{figure}

FP16 HMX tiles have a special memory layout, as shown in Figure~\ref{fig:hmx-fp16-crouton-tile-layout} (a). Both input and output tiles follow this layout. A typical way to construct this layout is to use HVX instructions to perform cross-lane shuffling on every two adjacent rows of the original matrix.

\subsection{Opportunities: Free Matrix Computation During LLM Decoding}

During the autoregressive generation process, LLM's input typically corresponds to only one token, which causes the GEMM operation to degenerate into GEMV. For example, an activation matrix of shape [1, \texttt{hidden\_dim}] is multiplied by a weight matrix of shape [\texttt{hidden\_dim}, \texttt{proj\_dim}]. In the case of using FP16 HMX, the effective size of each compute tile is [1,32]$\times$[32,32]. Since the basic unit of hardware computation is a $32\times32$ tile, 31 rows in the input activation tile do not correspond to actually useful content, resulting in low utilization of the matrix unit and waste of computing power.

\begin{figure}[h]
    \centering
    \includegraphics[width=0.8\linewidth]{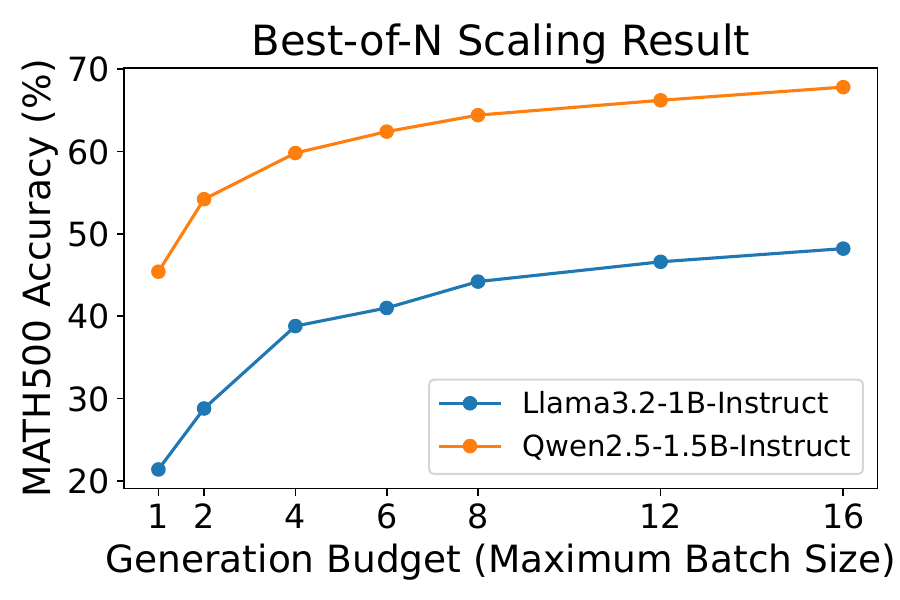}
    \caption{An example of test-time scaling with two models. The accuracy on MATH500 improves as generation budget increases.}
    \label{fig:tts-example}
\end{figure}

Meanwhile, some test-time scaling algorithms can achieve better generation quality by increasing the computation during generation, including parallel sampling methods such as Self-Consistency, Best-of-N and Beam Search. Their characteristic is that they explore multiple generation paths using a batch size greater than 1 and use certain ways (e.g., an external verifier) to select the better generation paths. Figure~\ref{fig:tts-example} shows an example of test-time scaling using Best-of-N. As the generation budget (i.e. the maximum batch size in the decoding phase) increases, the model accuracy on the MATH500 dataset improves significantly.

Based on these, we propose running the test-time scaling workloads of LLM on mobile NPUs. In this way, the computing power of the NPU wasted during the conventional LLM generation process can be effectively leveraged. The decoding overhead will not increase significantly in theory, and the generation quality of the model can be improved at run-time without modifying the model weights.

\subsection{Challenges}

Although it is theoretically feasible to utilize mobile NPUs for test-time scaling, an efficient implementation faces numerous hardware challenges. We summarize these challenges as follows.

\paragraph{Insufficient Precision.} Although HMX units support FP16 GEMM, deploying FP16 models on resource-constrained devices remains impractical, making quantized models the typical alternative. The matrix units in most mobile NPUs, including the HMX, were originally designed to accelerate integer-quantized DNN models that employ coarse-grained quantization schemes, such as per-tensor or per-channel quantization. As a concrete example, Hexagon NPUs lack native hardware support for the fine-grained quantization methods essential to modern LLMs. This limitation is further reflected in the software stack: QNN only supports per-tensor or per-channel weight quantization. Applying coarse-grained low-bit quantization directly to LLM weights can lead to significant accuracy degradation.

As shown in Table~\ref{tab:llama3.2-1b-performance-compare}, the accuracy results of the Llama 3.2 1B-Instruct model under QNN’s per-channel quantization\footnote{We use the official model released by PowerServe, available at https://huggingface.co/PowerServe/Llama-3.2-1B-PowerServe-QNN29-8G3} and AWQ per-group 4-bit quantization (both under W4A16 settings) indicate that the per-channel quantized model suffers severe performance degradation in challenging mathematical reasoning tasks. Unfortunately, since test-time scaling methods are applied in such tasks, the baseline accuracy achieved by QNN fails to meet even the minimal requirements for performance scaling.

\begin{table}[h]
    \centering
    \begin{tabular}{cccc}
        \hline
        dataset & AutoAWQ (W4A16) & QNN (W4A16) \\
        \hline
        MATH500 ($\uparrow$) & 15.9 & 2.1 \\
        GSM8K ($\uparrow$) & 32.6 & 3.4 \\
        Wiki PPL ($\downarrow$) & 19.42 & 28.99 \\
        \hline
    \end{tabular}
    \caption{Comparison of Llama3.2-1B-Instruct's performance under different implementation. QNN's quantization drastically hurts model's reasoning ability.}
    \label{tab:llama3.2-1b-performance-compare}
    \vspace{-1em}
\end{table}

\paragraph{Weak General Purpose Compute and Memory Bandwidth.} In the absence of native hardware support for fine-grained group quantization, a common approach is to rely on general-purpose computing units to handle such computations. However, we discover a significant gap between the compute and memory access capabilities of the general-purpose vector units and the specialized matrix units within the NPU. We measure the FP16 GEMM performance of both the HVX and the HMX on the Hexagon V75 NPU using a 1024$\times$1024$\times$1024 GEMM operation, with all inputs and outputs residing in on-chip TCM to reflect the hardware's peak performance. As shown in Table~\ref{tab:hexagon-npu-profile}, the FP16 GEMM throughput of the matrix unit reaches up to 12 TFLOPS — over 300 times higher than that of a single vector thread. In terms of memory bandwidth, a dedicated DMA engine achieves over 60 GB/s read bandwidth from DDR, whereas the vector unit's memory read bandwidth via the core data path remains below 30 GB/s. However, the high bandwidth provided by DMA is restricted to large, regular 1D or 2D data blocks and cannot efficiently handle small or irregular memory accesses. These observations highlight that the general-purpose compute and memory bandwidth of the vector unit are insufficient to keep up with the computational throughput of the specialized matrix unit, posing a major challenge for implementing high-performance mixed-precision GEMM kernels under fine-grained quantization.

\begin{table}[h]
    \centering
    \begin{tabular}{ccc}
        \hline
        hardware units & HVX (1 Thread) & HMX \\
        \hline
        FP16 GEMM GFLOPs & 32.93 & 12032.54 \\
        memory read bw. (GB/s) & 26 & 60 (DMA) \\
        \hline
    \end{tabular}
    \caption{The performance metrics of the HVX and HMX units, in terms of FP16 GEMM computing power (in GFLOPs) and memory read bandwidth (in GB/s).}
    \label{tab:hexagon-npu-profile}
    \vspace{-1em}
\end{table}

\section{Design Overview}

We present an LLM inference system designed for mobile NPUs and optimized for test-time scaling workloads.

To address the accuracy challenge, we adopt 4-bit fine-grained group quantization for the primary weights while keeping activations in floating-point. During runtime, we dynamically dequantize the weights into floating-point values on the fly, leveraging the powerful FP16 matrix computation capabilities of the NPU to efficiently support test-time scaling tasks.

For the unavoidable general-purpose computations — where the vector processing units exhibit limited memory bandwidth and compute throughput, our core strategy includes:

\begin{itemize}
    \item Employing hardware-aware offline design to minimize runtime computation overhead;
    \item Fully exploiting the intrinsic capabilities of SIMD vector units to bridge the gap between specialized hardware and flexible software requirements.
\end{itemize}

Specifically, we introduce the following techniques:

\paragraph{Hardware-aware Fine-grained Tile Quantization Scheme.} We propose a novel quantization layout that performs group quantization in fine-grained rectangular tiles, as opposed to conventional approaches that group along the accumulation axis. To align with the memory access patterns of both the matrix and vector units, we introduce an offline pipeline involving weight pre-quantization transformation, quantization, and post-quantization transformation. This enhances the continuity of runtime memory access and eliminates unnecessary computational overhead.

\paragraph{Efficient LUT-Based Computation.} For more complex runtime operations, we leverage the vector unit’s lookup table (LUT) instructions and generalized LUT mechanisms to replace intricate transformation logic. This approach accelerates key bottleneck operations in test-time scaling workloads, including dequantization within mixed-precision GEMM and the Softmax operation in Attention.

\section{System Design}

\subsection{Hardware-aware Fine-grained Tile Quantization Scheme}

Existing work~\cite{quantization-hurts-reasoning,quantization-meets-reasoning} has shown that quantization errors significantly degrade model performance in challenging tasks like mathematical reasoning. However, due to stringent on-device resource constraints, full-precision models remain infeasible, making fine-grained quantization essential to maintain accuracy.

Unfortunately, implementing efficient dequantization-based GEMM kernels under fine-grained quantization on mobile NPUs introduces substantial system challenges. We identify two primary issues:

\begin{itemize}
    \item mismatch between the weight layout expected by the matrix unit and conventional group quantization layout;
    \item suboptimal utilization of the wide vector registers caused by small group sizes.
\end{itemize}

To overcome these limitations, we propose a novel tile quantization scheme that incorporates two components:

\begin{itemize}
    \item a tile-based quantization strategy designed to align with the matrix unit's inherent data layout;
    \item a post-quantization weight permutation method that maximizes utilization of the vector unit's processing capabilities.
\end{itemize}

\subsubsection{Tile-Group Quantization}

In conventional quantized GEMM, weight matrices are typically stored in column-major layout, which aligns with the vector dot-product operations used in CPU-based matrix multiplication, such as in the llama.cpp CPU backend. The weights are divided into contiguous quantization groups — typically of size 32 — along the column dimension. Within each group, the values are quantized, and the resulting integer weights, along with their corresponding scale and zero-point parameters, are stored interleaved in memory, preserving the original column-wise ordering of the matrix.

\begin{figure}[h]
    \centering
    \includegraphics[width=0.8\linewidth]{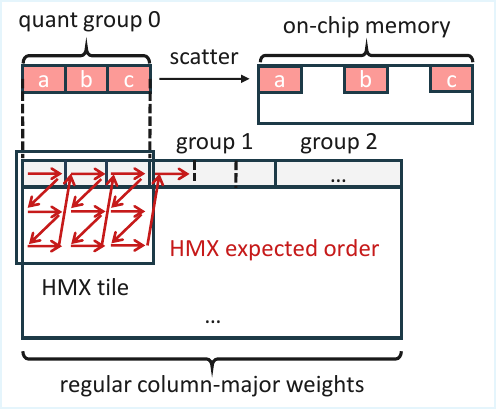}
    \caption{A simplified illustration of the mismatch between the quantization group layout and HMX tile layout.}
    \label{fig:dequant-layout-mismatch}
\end{figure}

However, on NPUs with special matrix units, the conventional group layout is often misaligned with hardware requirements. As illustrated in Figure~\ref{fig:dequant-layout-mismatch}, elements that are contiguous in the conventional layout become scattered in on-chip TCM. For SIMD vector units, such non-sequential access patterns are problematic. Although modern vector engines provide gather/scatter operations to alleviate scattered accesses, these operations remain expensive. Simply transposing the weight matrix does not resolve the mismatch, as the complex multi-level data layout expected by the matrix unit still results in noncontiguous memory access.

To address this, we first permute the weights into the layout expected by the matrix unit, and then apply round-to-nearest quantization group by group. For a group size of 32, this method effectively performs group quantization in units of $2\times16$ tiles. Given that pretrained weights in typical models approximately follow a zero-mean Gaussian distribution, quantizing within these reshaped tile groups does not significantly alter the statistical properties within each group compared to conventional grouping. Therefore, the resulting quantization error remains comparable.

Specifically, we arrange the weights before quantization according to the layout shown in Figure~\ref{fig:hmx-fp16-crouton-tile-layout}, which is hierarchically structured into two levels: an outer column-major ordering of tiles, matching the tile-level inner product operation of the matrix unit, and an inner shuffling of every two rows within each tile. We then quantize the weights group-wise in the new memory order.

\subsubsection{Coalescing Quantization Groups for Wide Vector Accesses}

By default, quantized weights are stored in an Array of Structures (AoS) layout. Taking Q4\_0 symmetric quantization as an example, each group of 32 elements consists of 16 bytes of INT4 quantized values and 2 bytes of FP16 scale values, with quantized values and scales interleaved in memory. Since memory access on the NPU architecture relies heavily on software-managed local 1D or 2D prefetching, we avoid the Structure of Arrays (SoA) layout, where quantized values and scales reside in separate large contiguous arrays, to better align with the hardware's preferred access pattern.

However, fine-grained quantization groups introduce a mismatch with the native vector processing granularity: A single quantization group is too small to fill a 128-byte wide vector register. Accessing such small groups would require multiple memory operations or additional instructions to merge data from multiple registers, resulting in inefficient memory bandwidth usage and computational overhead.

\begin{figure}
    \centering
    \includegraphics[width=0.7\linewidth]{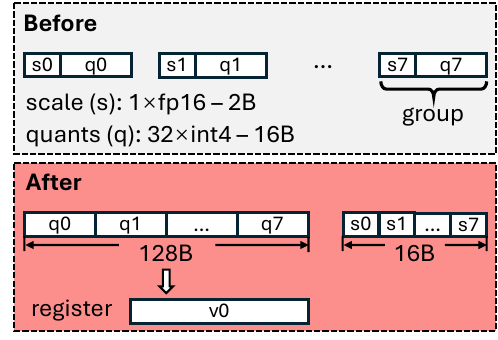}
    \caption{Repacking 8 fined-grained quantization groups into a super-block. The INT4 quantized values fit in a vector register.}
    \label{fig:dequant-group-repack}
\end{figure}

To solve the issue, we coalesce 8 quantization groups into a larger super-group and reorganize its content such that the INT4 values from 256 consecutive elements occupy exactly one full HVX register. This process is illustrated in Figure~\ref{fig:dequant-group-repack}.


\subsection{LUT-Based Computations}

Given the limited general-purpose computing performance of the vector unit, we propose using generalized look-up table (LUT) instructions to replace complex computations, thereby reducing instruction count and computational overhead. LUT-based computation is particularly effective for accelerating key operations in test-time scaling workloads, such as the exponential function in Softmax and the dequantization process.

\subsubsection{Fast Softmax via Vector Gather}

Test-time scaling methods typically increase sampling parallelism, leading to larger batch sizes and longer context lengths. We analyze the impact of these scaling factors on the major operators in transformer-based LLMs during generation:

\begin{itemize}
    \item GEMM. Based on previously described NPU hardware characteristics, moderately increasing batch size in test-time scaling workloads does not substantially increase GEMM latency. Moreover, GEMM latency is independent of context length.
    \item Misc. Ops. For operators such as activation functions, LayerNorm, residual Add, and RoPE, although their computational overhead is roughly proportional to input size, we neglect their impacts due to their small computation and memory access volumes.
    \item Attention. The theoretical computational complexity of Attention scales with both batch size and context length, making it a potential performance bottleneck in test-time scaling scenarios.
\end{itemize}

\begin{figure}
    \centering
    \includegraphics[width=0.8\linewidth]{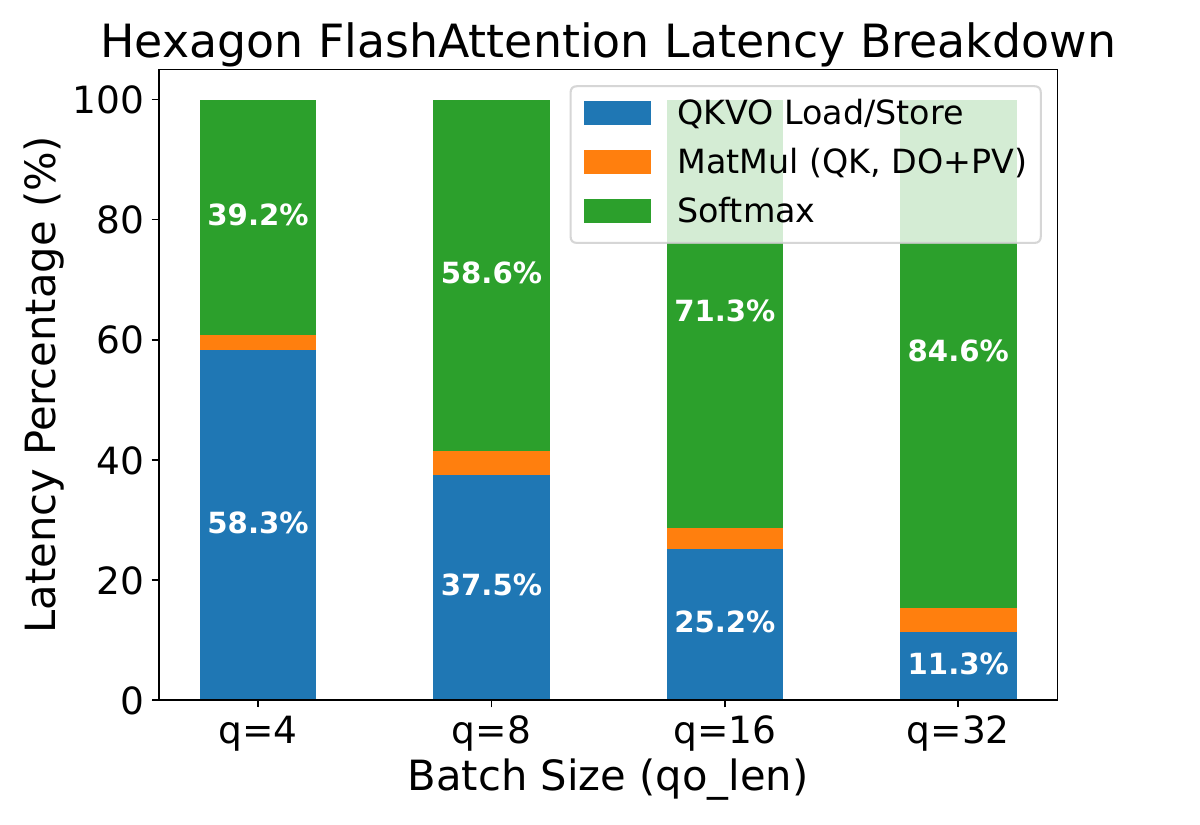}
    \caption{FlashAttention latency breakdown on Hexagon NPU. We use Qwen2.5-1.5B and prompt length is set to 4096.}
    \label{fig:fa-latency-breakdown}
\end{figure}

We implement FlashAttention~\cite{flash-attention2} on the Hexagon NPU using FP16 HMX and measure its latency composition at a prompt length of 4096 under various input batch sizes (query lengths), as shown in Figure~\ref{fig:fa-latency-breakdown}. The results indicate that matrix multiplication contributes little to overall latency, whereas Softmax dominates Attention execution time as the query length increases.

Our analysis shows that the primary bottleneck of on-chip Softmax lies in the exponential computation, which must be applied to $\Theta(N_q\times N_{kv})$ elements. Adding to the issue, these expensive exponential operations must be executed on the HVX, which lacks dedicated hardware support for special math functions. Following common practice, we replace \texttt{exp} with \texttt{exp2} and absorb the coefficient $\log_2{e}$ in the $QK^T$ scaling factor $\frac{1}{\sqrt{d}}$. For an input element $x$ decomposed into integer part $k$ and fractional part $f$, $2^f$ is approximated using a Taylor series polynomial expansion, while $k$ is directly added to the exponent field of $2^f$'s IEEE-754 representation. However, polynomial evaluation involves sequential dependencies, limiting instruction-level parallelism under the VLIW architecture.

To alleviate the exponential computation bottleneck, we explore replacing explicit exponent calculation with a precomputed lookup table (LUT). The HVX provides the \texttt{vgather} instruction, which can gather values from scattered locations in the TCM into a contiguous 128 byte TCM region. Although \texttt{vgather} can implement large LUTs, using LUTs for $\exp$ remains challenging: storing $2^{32}$ elements for 32-bit floats is impractical. Furthermore, \texttt{vgather} itself introduces substantial latency — 24 to 48 instruction packets on Hexagon V75, so its usage must be minimized.

To enable practical LUT-based $\exp$, we design the following approach. First, we extensively use FP16 throughout FlashAttention, with the on-chip computation process outlined in Algorithm~\ref{alg:fp16-fa-on-chip}. The matrices $S, P, O$ and the vectors $\vec m, \vec l$ are stored in 16-bit floats, with both the input and output of the $\exp$ computation in 16-bit floats. In particular, FP16 HMX uses higher-precision floating-point numbers for accumulation internally, and we upcast elements to 32-bit precision for critical operations such as row-wise summation of matrix $P$.

Using 16-bit inputs and outputs restricts the LUT to 65536 entries, requiring 128 KiB of storage, which fits within the TCM. A variant of \texttt{vgather} supports gathering 64 2-byte elements in one instruction, with a maximum address offset of 65536 bytes. However, 65536 FP16 entries occupy 128 KiB, leaving half of the entries inaccessible with direct addressing. To solve this, we leverage the property of safe softmax~\cite{online-softmax}, which ensures that all inputs to $exp$ are non-positive by subtracting the row-wise maximum $m_i$. Thus, we only store values for $x\le0$, resulting in a LUT with 32768 entries (64 KiB). During LUT-based $\exp$ computation, we ignore the MSB (sign bit) of the FP16 input and left-shift the input by one bit to generate the byte offset required by \texttt{vgather}.

The LUT is precomputed during system initialization, introducing no additional overhead during model inference. It occupies a fixed 64 KiB region in TCM, accounting for only $64KiB/8MiB\approx0.8\%$ of the total TCM capacity, thus minimally impacting TCM availability for other operations.

\begin{algorithm*}[h]
\caption{On-chip computation of ours FP16 FlashAttention (different heads omitted)}
\label{alg:fp16-fa-on-chip}
\KwIn{Head dimension $d$, Number of Query tiles $T_q$, Number of KV tiles $T_{kv}$, Query tile size $B_q$, KV tile size $B_{kv}$}
\KwIn{Matrices $Q_i$ (FP16) $\in \mathbb{R}^{B_q\times d}$, $K_j,V_j$ (FP16) $\in \mathbb{R}^{B_{kv}\times d}$}
Initialize $O_i^{(0)} = (\mathbf{0})\in \mathbb{R}^{B_q\times B_{kv}}$ (FP16), $m = (\mathbf{-\infty})\in\mathbb{R}^{B_q}$ (FP16), $l = (\mathbf{0})\in\mathbb{R}^{B_q}$ (FP16) \;
$S_i^{(j)} = \text{MatMul}(Q_i, K_j^T, \text{AccumType=\textcolor{blue}{FP32}}) \in \mathbb{R}^{B_q\times B_{lv}}$ (FP16) \;
$m_i^{(j)} = \max(m_i^{(j-1)}, \text{rowmax}(S_i^{(j)})) \in \mathbb{R}^{B_q}$ (FP16) \;
$P_i^{(j)} = \text{\textcolor{olive}{LUT\_Exp}}(S_i^{(j)} - m_i^{(j)}) \in \mathbb{R}^{B_q\times B_{kv}}$ (FP16) \;
$l_i^{(j)} = e^{m_i^{(j-1)}-m_i^{(j)}}l_i^{(j-1)} + \text{rowsum}(P_i^{(j)},\text{AccumType=\textcolor{blue}{FP32}}) \in \mathbb{R}^{B_q}$ (FP16) \;
$O_i^{(j)} = \mathrm{diag}(e^{m_i^{(j-1)}-m_i^{(j)}})O_i^{(j-1)}+P_i^{(j)}V_j \in \mathbb{R}^{B_q\times d}$ (FP16, AccumType=\textcolor{blue}{FP32}) \;
\KwOut{$O_i = \mathrm{diag}(l_i^{(T_{kv})})^{-1} O_i^{(T_{kv})}$}
\end{algorithm*}


\subsubsection{LUT-Centric Efficient Dequantization}

The runtime HVX dequantization requires careful design to avoid additional overhead. We present an efficient dequantization process based on the HVX lookup table instructions. The \texttt{vlut16} instruction is capable of performing a table lookup in a table of 16 elements for each 8 bit index in a source vector register. Each input byte is transformed into a 16-bit value, therefore \texttt{vlut16} results in a pair of registers.

\paragraph{Fast INT4 to FP16 conversion via table lookup}

\begin{figure}
    \centering
    \includegraphics[width=0.9\linewidth]{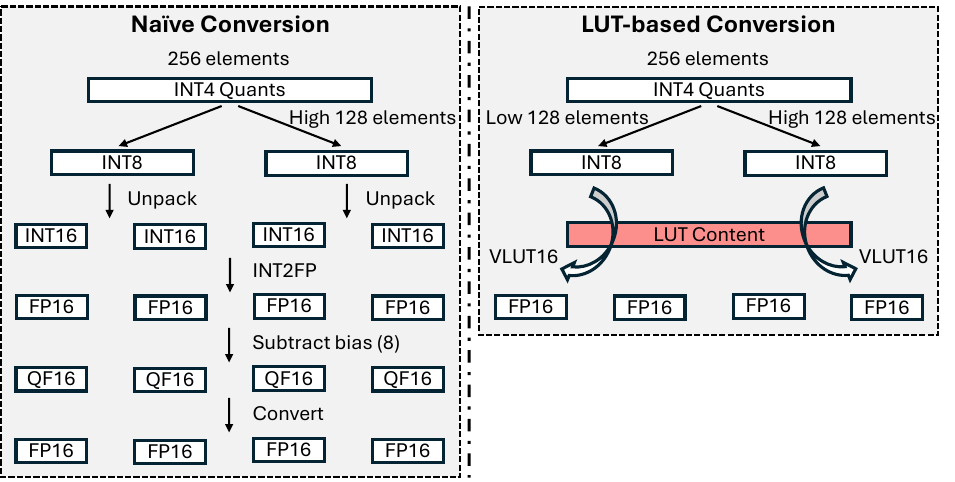}
    \caption{Converting INT4 quantized values into FP16 numbers via table lookup.}
    \label{fig:lut-based-int4-conversion}
\end{figure}

Using \texttt{vlut16} instructions, we directly transform 4-bits quantized values into [-8, 7] FP16 values for Q4\_0 quantization scheme, avoiding the conventional mask-unpack-convert instruction sequence. Figure~\ref{fig:lut-based-int4-conversion} demonstrates the comparison of two approaches. For Hexagon NPU prior to V79, all HVX floating-point operations produce results in an internal format called qfloat, which requires extra instructions to convert back to standard IEEE-754 formats. The use of table-lookup eliminates these overheads. This LUT-centric design can easily support different 4-bit encoding schemes (e.g. FP4, NF4~\cite{qlora}, IQ4\_NL used in llama.cpp) simply by adjusting the table contents.

\paragraph{Scales broadcast via table lookup}

A 128-byte HVX register can accommodate two FP16 quantization groups of size 32. Therefore, the conventional approach is to broadcast scalar scales to the entire vector register and then concatenate two registers for subsequent multiplication with quantized values. However, by using the scales of four groups as LUT contents and applying predefined constant indices, we can achieve the broadcast of four groups of scales with just one \texttt{vlut16} instruction.







\section{Implementation}

Our inference system is implemented on top of llama.cpp~\cite{llama.cpp} with approximately 7K lines of code in C/C++ and inline assembly. We use the LLVM toolchain in the Hexagon SDK (version 6.0.0.2) to generate code for Hexagon NPUs. We especially note that our system has no dependency on Qualcomm's QNN, avoiding inflexible static fixed-shape computation graphs.

Our implementation mainly consists of two modules: one module is the operator library for the Hexagon NPU, which is compiled into an independent Hexagon DSP shared object; the other module is integrated with llama.cpp on the CPU side. The NPU operator library implements computation kernels, power management, hardware resource management, and a computation thread pool. We add a Hexagon NPU backend to llama.cpp, leveraging \texttt{rpcmem} shared memory as the underlying buffer type. \texttt{rpcmem} is a wrapper for the kernel dmabuf memory and supports the sharing of physical memory between the CPU and the NPU. The related allocation, deallocation, and mapping interfaces are provided by \texttt{libcdsprpc.so} in the Android system’s vendor libraries. By utilizing shared memory buffers, we not only eliminate unnecessary inter-processor data copy but also reuse the existing memory management system as much as possible. In addition, we are able to schedule the operators that have not been implemented on the NPU to run on the CPU, achieving seamless integrations with upper-layer applications.

During the backend initialization phase, we call the FastRPC~\cite{fastrpc} facility of the Hexagon SDK to start the remote NPU session and initialize an area of shared memory for communication. On the NPU side, a thread continuously polls in this shared-memory area to receive computation requests from the CPU. Compared to the default RPC implementation, communication through shared memory can have a lower latency. We note that after the CPU writes data to the shared memory, the NPU will not automatically invalidate the cache of the corresponding area as there is only one-way coherence between the CPU and the NPU on the Snapdragon SoC. Therefore, we manually clear the cache before NPU polls. Similar cache maintenance operations are also required for shared buffers containing model activations.

\section{Evaluation}





\subsection{Experiment Setup}

\textbf{Devices.} The experiments on NPU performance are conducted on three Android devices: OnePlus Ace3, OnePlus 12, OnePlus Ace5 Pro. Some of the accuracy results are obtained on a server testbed equipped with NVIDIA RTX3090 GPUs.

\begin{table}[]
    \centering
    \begin{tabular}{ccc}
        \hline
        Device & SoC & NPU Arch.\\
        \hline
        OnePlus Ace3 & Snapdragon 8 Gen 2 & V73 \\
        OnePlus 12 & Snapdragon 8 Gen 3 & V75 \\
        OnePlus Ace5 Pro & Snapdragon 8 Elite & V79 \\
        \hline
    \end{tabular}
    \caption{Mobile devices used in evaluation.}
    \label{tab:mobile-devices}
    \vspace{-1em}
\end{table}

\textbf{Models.} We choose models from the Qwen 2.5~\cite{qwen25} and Llama 3.2~\cite{llama32} model family. Considering the actual resource limitations of mobile phones, we mainly evaluate Qwen 2.5 with model sizes of 1.5B and 3B, as well as Llama 3.2 with model sizes of 1B and 3B, which correspond to practical deployable model sizes. When evaluating the performance-cost trade-off of time-time scaling methods, we additionally consider Qwen 2.5 with a model size of 7B. In the evaluation of mathematical reasoning tasks, we use the Instruct model variants of Qwen 2.5 and Llama 3.2. For Best-of-N search and step-level beam search, Skywork-1.5B-PRM~\cite{Skywork-o1} is used as the outcome-reward and process-reward scorer.

\textbf{Datasets and metrics.} In the test-time scaling tasks, we evaluate the pass@1 accuracy of the models in two mathematical reasoning tasks, MATH500~\cite{math} and GSM8K~\cite{gsm8k}, and we uniformly use the 0-shot CoT prompt. For other accuracy measurements, the WinoGrande~\cite{winogrande} accuracy, the MMLU~\cite{mmlu} accuracy, and the Wikitext-2 perplexity are evaluated using \texttt{llama-perplexity} utility.

\textbf{Baselines.} Since we focus on test-time scaling tasks, we mainly present the performance of our implementation under different decoding workloads. To demonstrate the advantages of using NPUs to run test-time scaling workloads, we select the recent OpenCL backend of llama.cpp\footnote{commit: 1caae7f} as the GPU-based system for comparison. This OpenCL backend incorporates optimized Q4\_0 matrix multiplication kernels tailored for Snapdragon's Adreno GPU. Since existing NPU-based systems all have certain limitations in handling test-time scaling workloads, we do not use them as the primary baselines: \texttt{llm.npu}~\cite{mllm-npu} does not utilize the NPU for computation during the decoding phase; other QNN-based systems have low accuracy (e.g., PowerServe~\cite{powerserve}); and systems like Powerinfer-2~\cite{powerinfer2} and HeteroLLM~\cite{heterollm} are not open-source. Nevertheless, we still report the QNN-based data as a reference in Section~\ref{sec:system-comparison}.


\textbf{Settings.} In the operator-level evaluation of GEMM, we select the sizes of the weight matrices of the linear layers corresponding to Qwen2.5-1.5B, Qwen2.5-3B, Llama3.2-1B, and Llama3.2-3B. Specifically, these include the Attention projection matrices $W_q,W_o$ and $W_{gate}, W_{up}, W_{down}$ in the Feed Forward Network (FFN). (For modern models that use Grouped Query Attention (GQA), the projection matrices $W_k,W_v$ in Attention are not selected because their scale is smaller compared to $W_q,W_o$). Most of the matrices adopt the Q4\_0 quantization scheme, which corresponds to 4.5 Bits Per Weight (BPW). As for the FFN down matrices, we apply the Q8\_0 quantization scheme (8.5 BPW) to reduce quantization errors, as existing work indicates their importance in preserving model accuracy~\cite{kovaleva2024systematic,duquant,lin2023emergent}.

\subsection{Overall Performance}

\subsubsection{Accuracy-Latency Trade-off of Test-time Scaling}

\begin{figure*}
    \centering
    \includegraphics[width=1.0\linewidth]{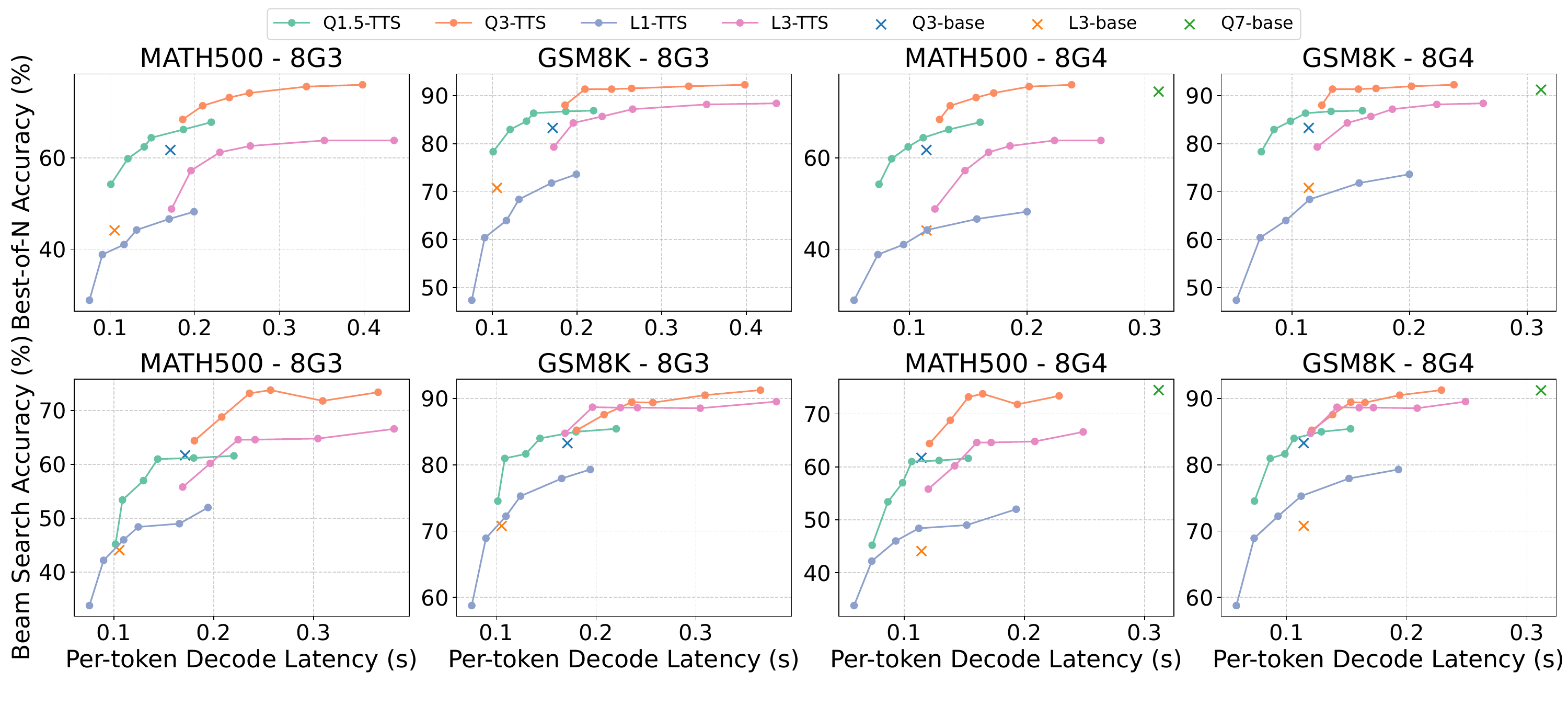}
    \caption{Accuracy-latency trade-off of different test-time scaling methods on various combinations of dataset and hardware.}
    \label{fig:accuracy-latency-tradeoff}
\end{figure*}

Figure~\ref{fig:accuracy-latency-tradeoff} illustrates the performance-cost trade-off of the test-time scaling methods. We use the accuracy in MATH500 and GSM8K as metrics for generation quality and the average decoding latency of on-device models as the cost metric (the data here account for the increased context length introduced by TTS). In the figure, the top row and the bottom row correspond to Best-of-N and Beam Search results, respectively, while "QN"/"LN" denotes the Qwen2.5 or Llama3.2 models with $N$ billion parameters. The SoC results exclude the "8G2" entry due to a known NPU virtual address space limitation~\cite{fastrpc-issue-137} of Snapdragon 8 Gen 2 that prevents models with 3B or more parameters from running. The isolated points marked with a "base" represent the average performance obtained via conventional sampling with the models.

The data show that test-time scaling offers a trade-off space and achieves a more superior Pareto frontier under specific configurations, enabling a better performance-cost balance. In the Best-of-N method, the scaling results of Qwen2.5 1.5B and 3B outperform the baseline accuracies of the 3B and 7B models, respectively. For Beam Search, Qwen2.5-1.5B and Llama3.2-1B can achieve efficiency comparable to or slightly better than their respective 3B variants. Our results indicate that by leveraging the computing power of NPUs and test-time scaling algorithms, small on-device models have the potential to surpass larger models in terms of both generation quality and inference cost.

\subsubsection{On-Device Decoding Performance}

\begin{figure*}[h]
    \centering
    \includegraphics[width=0.7\linewidth]{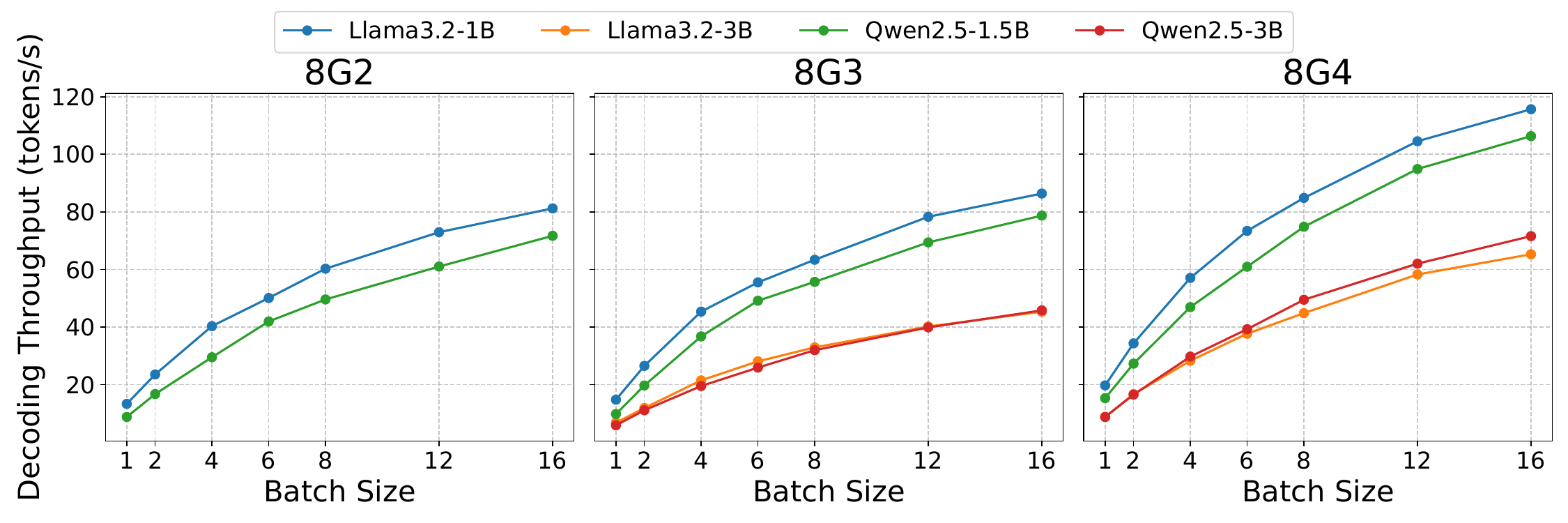}
    \caption{End-to-End decoding throughput of different models under various batch sizes and hardware settings.}
    \label{fig:decoding-throughput}
\end{figure*}

Figure~\ref{fig:decoding-throughput} demonstrates the on-device decoding throughput of different models in different batch sizes. We only evaluate Qwen2.5-1.5B and Llama-3.2-1B on OnePlus Ace3 due to a 2GiB limitation of the virtual address space on older NPUs.

The data show that for the three devices, the end-to-end decoding throughput of the system significantly increases as the batch size increases. The fundamental reason for the increase in decoding throughput is that the idle computing power of the HMX unit is utilized, and, essentially, the computation time consumed on the core HMX does not increase at all. However, the decoding throughput does not scale perfectly linearly because the inference process contains parts that become much slower with the growth of the input length. Specifically, in our implementation, we conservatively place the weights of the \texttt{lm\_head} (the projection matrix from the hidden states to the vocabulary) and the related activations on the CPU instead of the NPU. Modern LLMs have a large vocabulary, making the \texttt{lm\_head} and logits occupy a large space. Unfortunately, the Hexagon NPU only has a 32-bit virtual address space, therefore placing the complete logits tensor on the NPU may prevent the complete model from running. Currently, we observe that when the batch size equals 16, the proportion of the computation time of logits on the CPU is close to or exceeds 50\%. We expect that after addressing the limitations of the NPU address space and placing the logits computation on the NPU, the system will achieve better throughput scaling characteristics.

\subsubsection{Power and Energy Consumption}

\begin{figure}[h]
    \centering
    \includegraphics[width=1.0\linewidth]{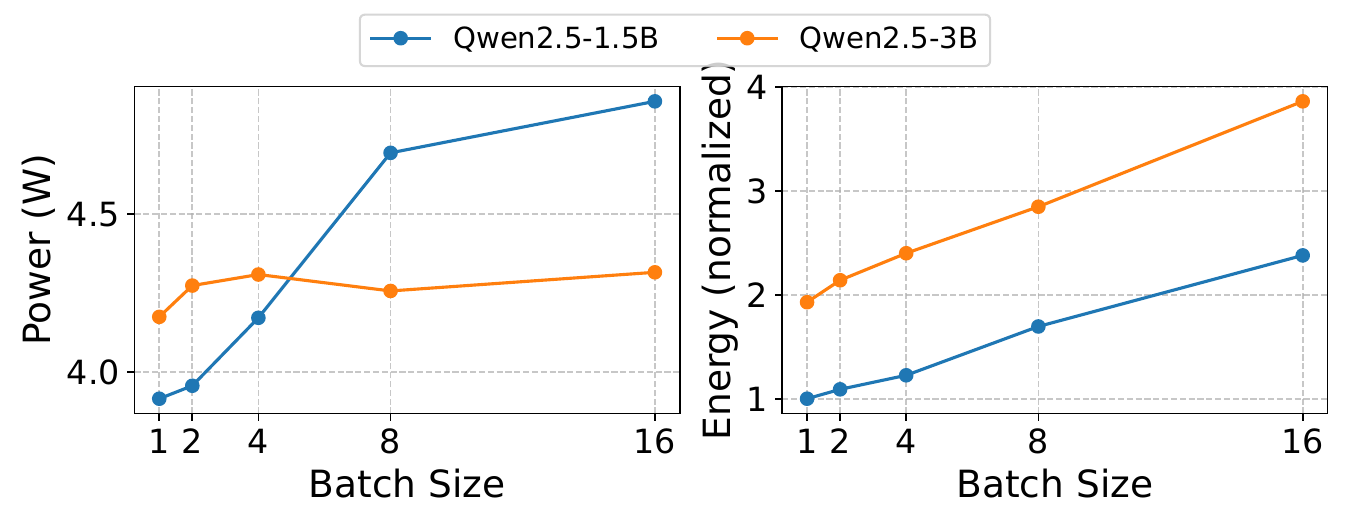}
    \caption{Power and energy consumption during the LLM decoding stage.}
    \label{fig:power-consumption}
\end{figure}

We measure the power consumption during LLM decoding via sysfs interface on OnePlus 12 with the performance mode enabled. As the batch size increases in the decoding phase, the power consumption of running the 1.5B Qwen model increases, but the overall power consumption of the device is still within 5W; in contrast, the power consumption corresponding to running the 3B Qwen model stabilizes at around 4.3W. Figure~\ref{fig:power-consumption} shows the normalized energy consumption, which is calculated by multiplying the corresponding power consumption by the relative decoding latency. The scaling trait of energy consumption with respect to the batch size is similar to that of decoding latency; therefore, replacing the cost metric in Figure~\ref{fig:accuracy-latency-tradeoff} with energy also results in similar accuracy-cost trade-off characteristics. In particular, we note that the decoding energy consumption of the 1.5B model at a batch size of 8 is lower than that of the 3B model at a batch size of 1, while the test-time scaling accuracy of the 1.5B model when decoding with a batch size of 8 on mathematical tasks is comparable to the base accuracy of the 3B model.

\subsubsection{Comparison with Other Systems} \label{sec:system-comparison}

\begin{figure}[h]
    \centering
    \includegraphics[width=1.0\linewidth]{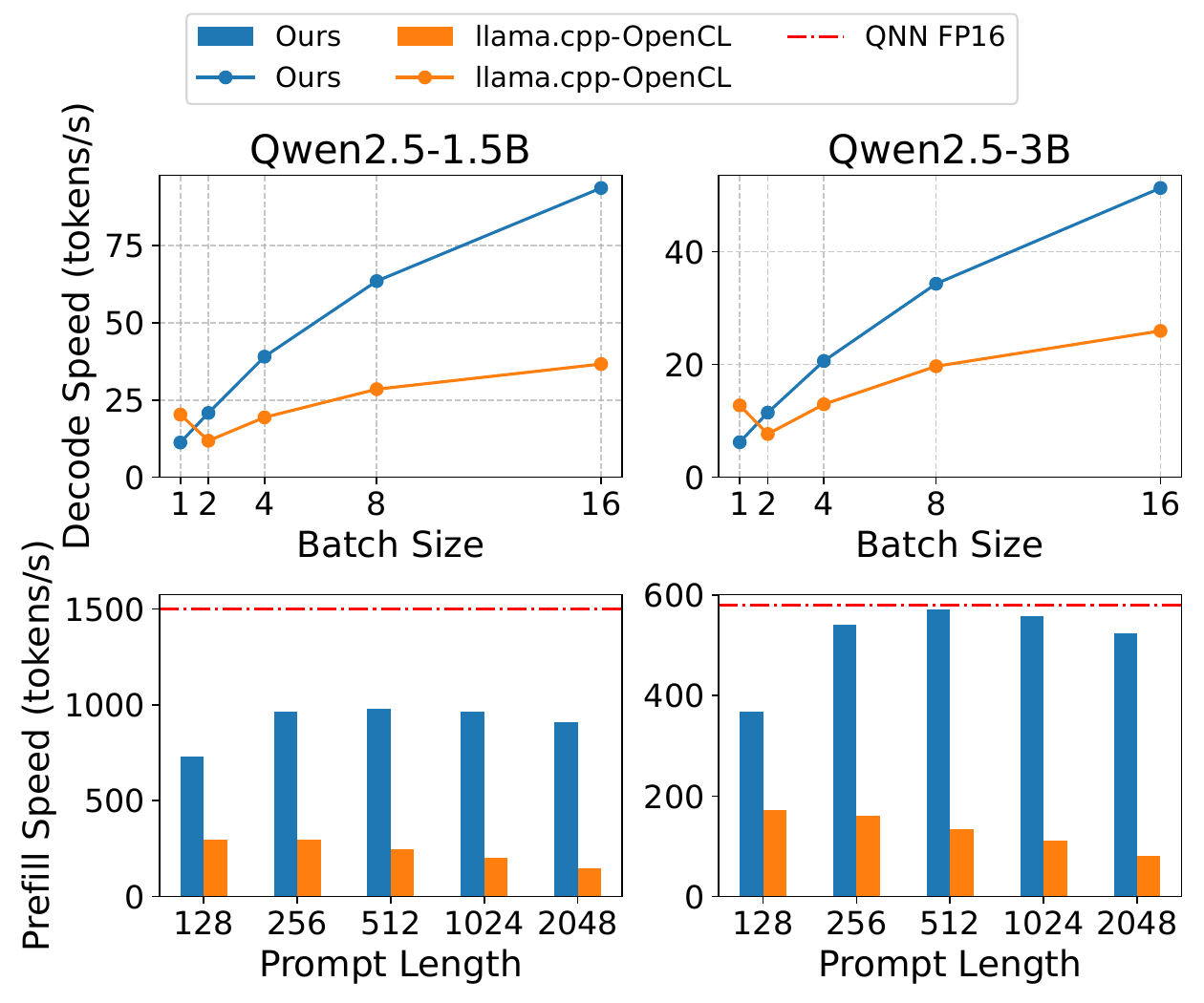}
    \caption{Inference throughput comparison.}
    \label{fig:system-comparison}
\end{figure}

The decoding and prefilling performance of our system is presented in Figure~\ref{fig:system-comparison}. We compare our system against a GPU-based implementation and add the performance of FP16 QNN as a reference. During the decoding phase, although the GPU decodes faster at batch size 1, our NPU-based system exhibits higher decoding throughput and better scaling characteristics at larger batch sizes, highlighting the advantage of using NPUs in test-time scaling workloads. Our system also consistently outperforms the GPU-based system in terms of prefilling throughput, achieving comparable performance with proprietary QNN under certain workloads.

\subsection{Accuracy Assessment}

\begin{table}[h]
    \centering
    \begin{tabular}{cccc}
        \hline
        dataset & Tile group & Common group & F16 \\
        \hline
        WinoGrande ($\uparrow$) & 62.559 & 63.349 & 64.613 \\
        MMLU ($\uparrow$) & 35.465 & 35.271 & 34.819 \\
        Wiki PPL ($\downarrow$) & 10.206 & 10.190 & 9.798 \\
        \hline
    \end{tabular}
    \caption{Accuracy comparison between models using tile quantization groups tailored for HMX layout and models using conventional quantization groups.}
    \label{tab:quantization-group-accuracy-compare}
    \vspace{-1em}
\end{table}

\paragraph{Quantization Scheme.} We evaluate the accuracies of the Qwen2.5-1.5B model corresponding to the tile quantization groups based on the HMX layout and the conventional quantization groups. As shown in Table~\ref{tab:quantization-group-accuracy-compare}, the model using our quantization layout has slightly higher accuracy in MMLU compared to the model with the conventional layout, and there is only a slight decrease in Winogrande and Wikitext PPL. Moreover, these accuracy differences are much smaller than the performance loss caused by quantization itself (as indicated by Wikitext perplexity in the "F16" column). In general, using our proposed tile quantization group does not lead to a significant decrease in the accuracy of the quantized model.

\begin{table}[h]
    \centering
    \begin{tabular}{cccc}
        \hline
        dataset & Our LUT16 FA & F32 Attention  \\
        \hline
        WinoGrande ($\uparrow$) & 62.796 & 62.559 \\
        MMLU ($\uparrow$) & 35.207 & 35.465 \\
        Wiki PPL ($\downarrow$) & 10.205 & 10.206 \\
        \hline
    \end{tabular}
    \caption{Accuracy comparison between models using our F16 FlashAttention with LUT-based Softmax and models using conventional F32 Attention.}
    \label{tab:fa-method-accuracy-compare}
    \vspace{-1em}
\end{table}

\paragraph{Attention Implementation.} Table~\ref{tab:fa-method-accuracy-compare} shows the model accuracies corresponding to our LUT-based FP16 Attention and the conventional FP32 Attention, using the same model and datasets as above. It can be seen that replacing the non-critical parts in Attention (except for the accumulation) with a lower FP16 precision does not have a noticeable impact on the end-to-end accuracy of the model.

\subsection{Ablation Study}

\begin{figure}[htbp]
    \centering
    \includegraphics[width=1.0\linewidth]{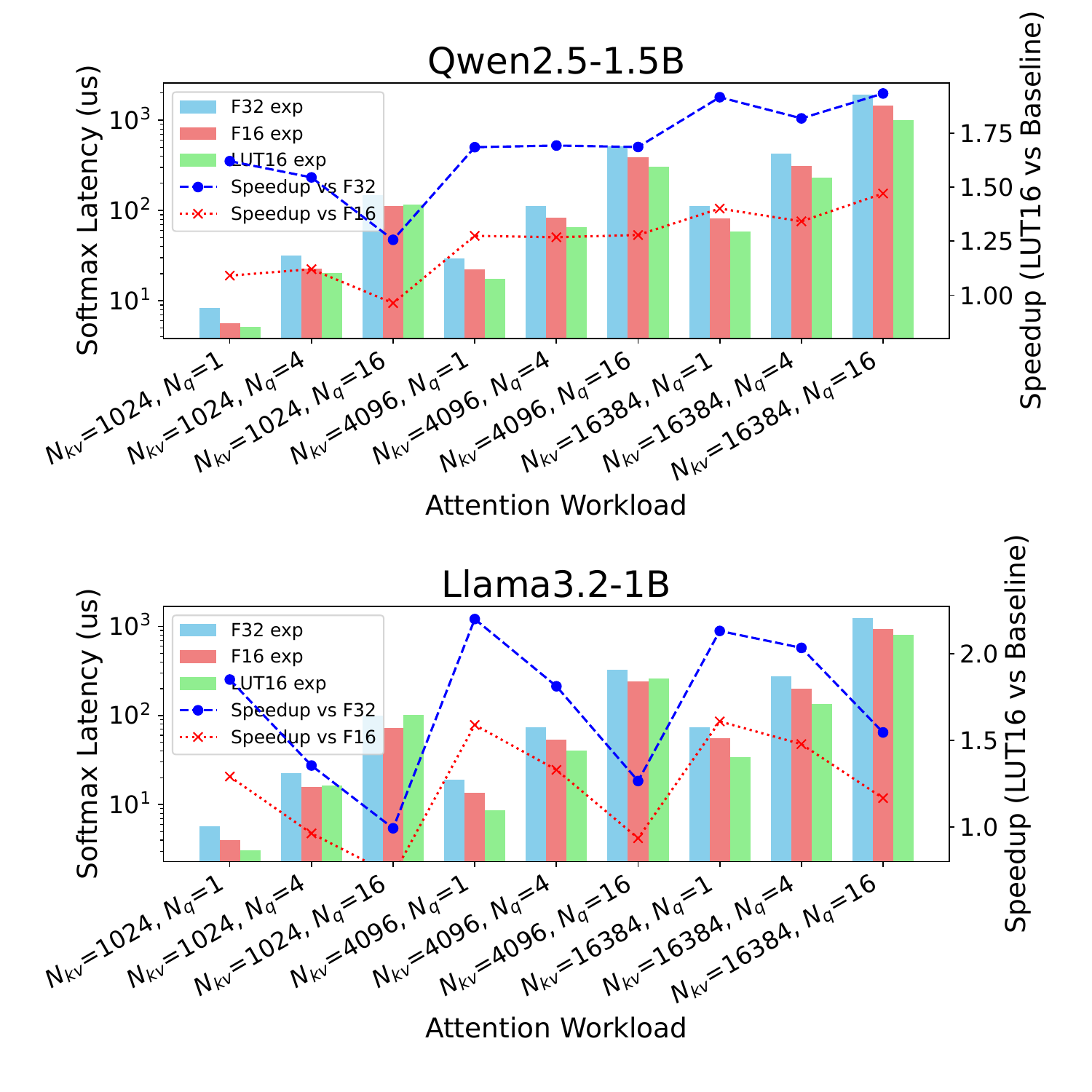}
    \caption{Ablation study of on-chip softmax of our proposed F16 Attention with LUT-based exponential computation. Performance is measured on OnePlus 12.}
    \label{fig:softmax-exp-compare}
\end{figure}

\paragraph{Softmax in Attention.} Figure~\ref{fig:softmax-exp-compare} shows the on-chip softmax latency corresponding to the calculation of the exponential function $\exp$ using different methods under different attention workloads. The length of the input query for Attention is set to 1, 4, and 16, while the length of KV is set to 1024, 4096, or 16384. The figure indicates that our LUT-based exponential calculation achieves an acceleration of 1.26 to 2.19 times compared to the conventional 32-bit floating-point $\exp$, and up to $1.60\times$ speedup compared to the 16-bit floating-point $\exp$. It is worth noting that when pre-computing the $\exp$ lookup table, floating-point numbers with a width of 32 bits or higher can be used to calculate the intermediate results. Therefore, the LUT-based $\exp$ has a higher accuracy than the 16-bit polynomial approximation of $\exp$. When the context length is short, a larger input query will slightly reduce the acceleration ratio, but this phenomenon will be alleviated when the KV length is longer.

\begin{figure*}[htbp]
    \centering
    \includegraphics[width=1.0\linewidth]{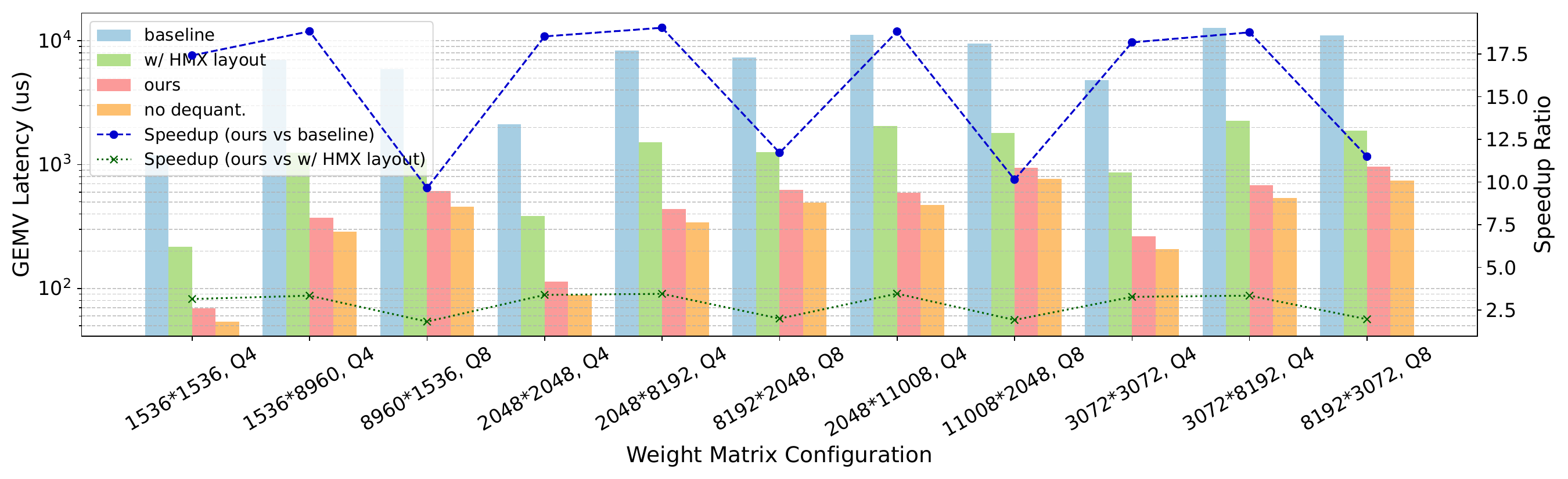}
    \caption{Ablation study of proposed optimizations on GEMM dequantization. We measure the performance of GEMV on OnePlus 12.}
    \label{fig:gemv-latency-ablation}
\end{figure*}

\paragraph{Dequantization-based GEMM} Figure~\ref{fig:gemv-latency-ablation} presents the ablation experiment for optimization of the GEMM dequantization layout. The baseline method corresponds to the conventional memory layout, where the column-major weight matrix is quantized according to the continuous groups in memory. The GEMM kernel dequantizes the 32-sized groups one by one during runtime and then scatters the elements to the correct positions in the TCM. The item of "HMX layout" applies the offline weight rearrangement and tile quantization group for the HMX layout, enabling the FP16 weights to be continuously written into the TCM. "Ours" is the version that adopts all the optimizations including the quantization group coalesce. In addition, we add a set of data labeled "no dequantization". In this implementation, instead of performing actual weight dequantization, the quantized weights are read directly from the memory and copied to the on-chip memory without any computation. This set of data can be regarded as the performance upper bound of dequantization-based methods.

Compared to the baseline, our method achieves an acceleration of 9.65 to 19.04 times under different matrix sizes. This is mainly because the scatter operations in the baseline are extremely costly. After applying the HMX layout, the quantization group coalesces and the rearrangements also effectively reduce computational waste, bringing a speedup of $1.82\times$ to $3.45\times$. In particular, compared to the "no dequantization" group, our method is only 27\% slower on average, indicating that this implementation is already close to the performance upper limit of dequantization.

\subsection{Overhead and Sensitivity Analysis}

\begin{figure}[h]
    \centering
    \includegraphics[width=1.0\linewidth]{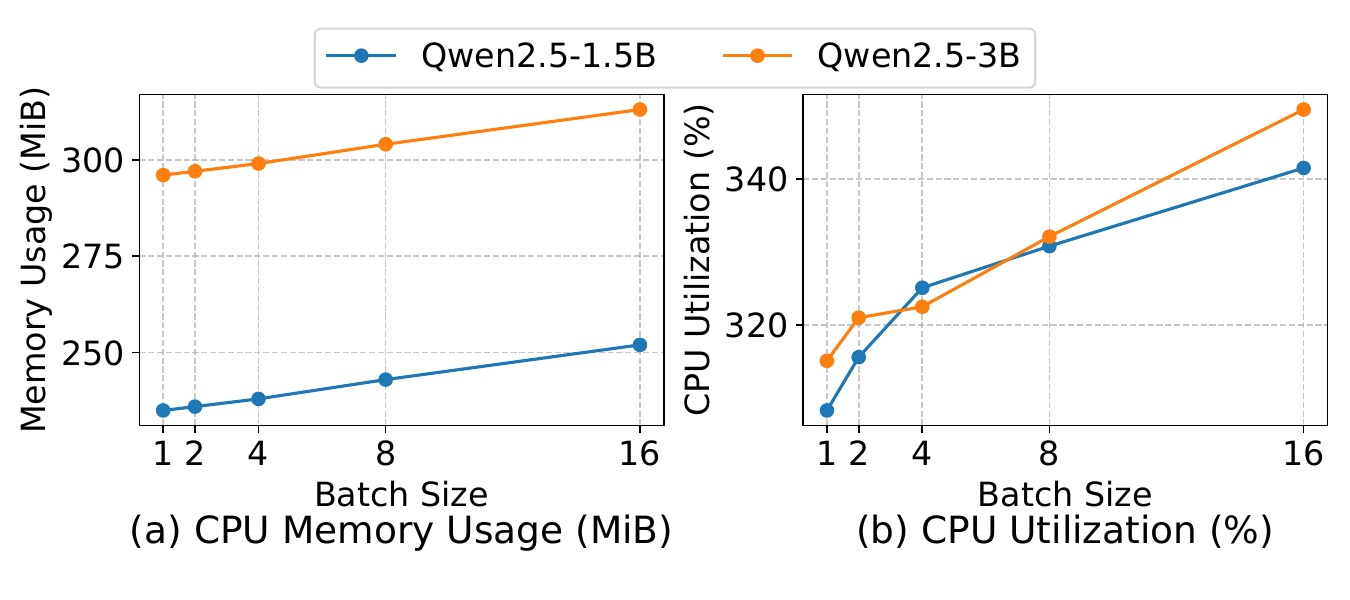}
    \caption{CPU and memory usage during the decoding stage.}
    \label{fig:cpu-memory-usage}
\end{figure}

\paragraph{CPU and Memory Usage.} We evaluate the CPU utilization and memory consumption of the 1.5B and 3B Qwen2.5 models during the decoding stage on OnePlus 12. The CPU memory usage presented in Figure~\ref{fig:cpu-memory-usage} is derived from the resident memory size reported by the \texttt{top} command. We also measure the total size of \texttt{dmabuf} (i.e., memory used by NPU) using \texttt{pmap}, yielding constant values of 1056 MiB and 2090 MiB under a context budget of 4096 tokens for the 1.5B and 3B models, respectively. The total memory consumption is approximately 1.3 GiB for the 1.5B model and 2.4 GiB for the 3B model. The CPU utilization increases with batch size due to the increased computation of vocabulary projection on CPU, yet the number of utilized cores is consistently limited to 4.

\begin{figure}[h]
    \centering
    \includegraphics[width=1.0\linewidth]{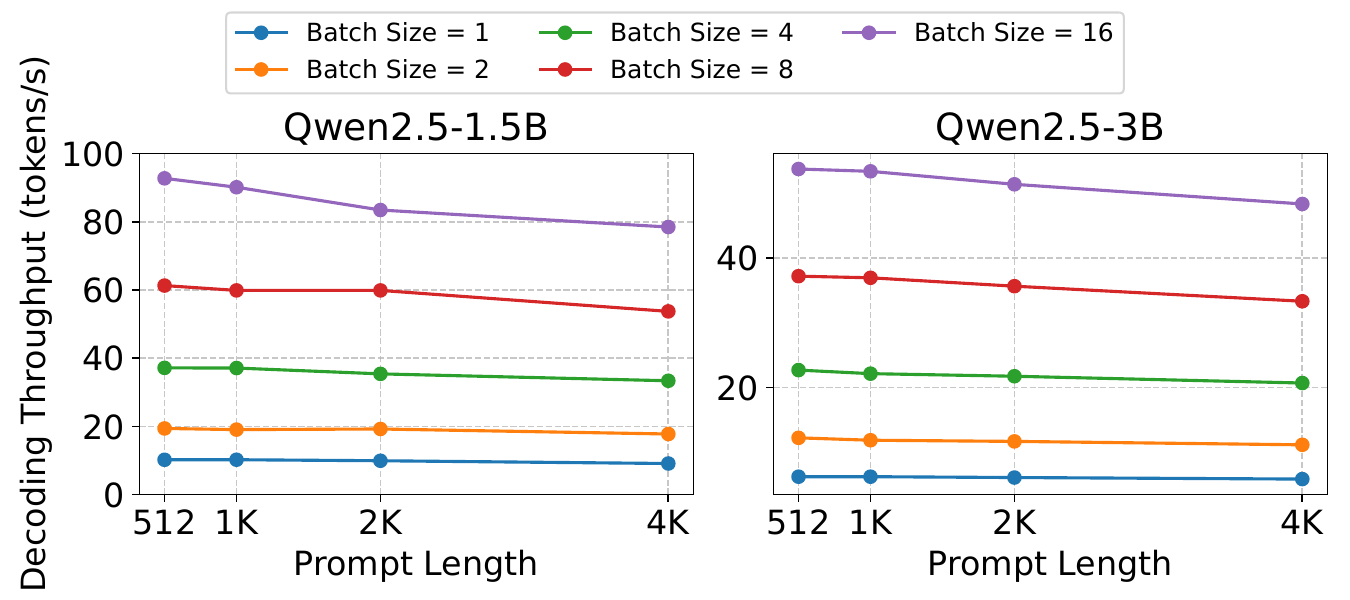}
    \caption{Impact of prompt length on decoding throughput.}
    \label{fig:prompt-length-sensitivity}
\end{figure}

\paragraph{Impact of Prompt Lengths.} Figure~\ref{fig:prompt-length-sensitivity} shows the impact of prompt lengths on decoding throughput. Across all batch sizes and both models, the decoding throughput exhibits a mild decreasing trend as the prompt length increases from 512 to 4096 tokens. However, within the range of prompt lengths up to 4096 tokens, this decline remains relatively subtle, indicating that prompt length exerts only a limited influence on decoding throughput in this interval.

\section{Discussion}

\paragraph{Generalizability to Other Hardwares.} We argue that the “vector + matrix” architecture of NPUs possesses a certain degree of universality and observe that the boundary between CPUs and NPUs is gradually blurring. Beyond NPUs, modern CPUs have also begun to incorporate dedicated matrix multiplication units, such as Intel AMX and ARM SME, endowing them with a similar "vector + matrix" architecture. Furthermore, we note that modern AI accelerators generally exhibit a significant disparity between general-purpose computing performance and specialized low-precision matrix multiplication capabilities (e.g. NVIDIA GPUs). Although specific hardware architectures may differ, the core ideas behind our techniques maintain broad applicability.

\paragraph{System Performance and Limitations.} (a) Decoding Performance: The current decoding speed of our system is relatively constrained, primarily due to the overhead of dequantization. However, this does not undermine the effectiveness of test-time scaling. Quantized GEMM based on QNN typically utilizes only the DMA and HMX components without introducing HVX computational overhead. Approaches similar to T-MAC~\cite{tmac} could potentially enable efficient GEMV with fine-grained group quantization on NPUs, thereby accelerating the LLM decoding process. (b) Prefill Performance: There remains room for improvement in the prefill performance of our current system. Offloading more operators to the NPU, reducing memory access and communication overhead through operator fusion, and optimizing tiling and pipelining strategies for matrix multiplication could all contribute to enhanced prefill performance. We leave these optimizations to future work. (c) Model Size Constraints: Our current implementation is limited by the 32-bit address space of a single NPU session on older devices. Employing multiple NPU sessions could help alleviate this issue.

\paragraph{Application Scope of Parallel Test-time Scaling.} Although parallel test-time scaling methods currently dominate mathematical reasoning tasks, evidence from recent studies~\cite{yan2025empowering,tot,paranjape2023boosting,hong2023metagpt,hafner2024reasoning,parallel-scaling-law} indicates their extensibility to broader reasoning and planning domains, highlighting substantial generalizable potential.

\section{Related Works}


\paragraph{On-Device LLM Inference with NPUs.} \texttt{llm.npu}~\cite{mllm-npu} pioneered the use of per-tensor quantized INT8 GEMM on NPUs to accelerate the prefill phase of LLMs, employing the CPU to assist in outlier-related computations to maintain accuracy. HeteroLLM~\cite{heterollm} achieves collaborative inference between GPUs and NPUs through tensor partitioning. PowerServe~\cite{powerserve}, an open-source inference framework, leverages the intermediate ONNX format to implement custom quantized and floating-point computation partitioning. ShadowAttn~\cite{shadowattn} utilizes collaboration between NPUs and CPUs/GPUs to accelerate sparse attention. ExecuTorch~\cite{executorch,torchao} is a well-known open-source edge-side DNN inference framework, supporting SpinQuant~\cite{spinquant} and the QNN backend. All of the above works are based on Qualcomm’s Hexagon NPU and use the closed-source QNN~\cite{qualcomm-qnn} as the backend. Works from vivo~\cite{bluelm,edgeinfinite} have utilized MediaTek’s NPU, but due to the non-public accessibility of the NeuroPilot SDK, such research remains scarce.

\paragraph{LLM Quantization.} The most well-known post-training quantization algorithms include GPTQ~\cite{gptq} and AWQ~\cite{awq}, they perform weight-only quantization and only require small amounts of calibration data, therefore they are extensively used. Subsequent methods such as SmoothQuant~\cite{smoothquant}, DuQuant~\cite{duquant} broadened the scope to include weight-activation quantization, tackling the more challenging task of quantizing activations by developing techniques to mitigate their problematic outlier distributions. Most recently, comprehensive approaches such as QuaRot~\cite{quarot} and SpinQuant~\cite{spinquant} have emerged, aiming to quantize all major components - weights, activations, and the critical KV cache - often down to 4-bits. These methods leverage rotation transformations to create more quantization-friendly feature distributions throughout the model.

\paragraph{Speculative Decoding.} Speculative Decoding~\cite{spec-decode,specinfer,zengyixiao-sd} is a class of acceleration methods for LLM inference, the core of which is to verify multiple speculated tokens in one model forward pass to alleviate the memory-bound issue of LLM decoding. There are various extended variants of Speculative Decoding, and some~\cite{hao2024hybrid,reward-guided-sd} no longer strictly follow the distribution of the target model. In theory, generalized Speculative Decoding and test-time scaling methods both belong to the generalized Generate-then-Verify framework, and our system can theoretically support these applications seamlessly.

\section{Conclusion}

This work demonstrates the feasibility and effectiveness of leveraging the underutilized compute capacity of mobile NPUs — specifically the Qualcomm Hexagon NPU — for test-time scaling of LLMs. By designing an end-to-end inference system incorporating hardware-aware tile quantization, weight layout optimization, and LUT-based acceleration of key operators, we show that smaller models augmented with test-time scaling can outperform larger conventionally-deployed models in both accuracy and latency. This approach provides a new pathway for deploying high-performance language models on resource-constrained mobile devices, advancing the Pareto frontier of efficiency and capability for on-device AI.


\section{Acknowledgement}

We thank all the anonymous reviewers and our shepherd, Yubin Xia, for their insightful feedback and suggestions. This work was supported by Tsinghua University (AIR)-AsiaInfo Technologies (China), Inc. Joint Research Center for 6G Network and Intelligent Computing under Grant 20233910006.

\balance
\bibliographystyle{plain} 
\bibliography{refs}


\end{document}